\documentclass[prd,aps,floats,floatfix,eqsecnum,nofootinbib]{revtex4}
\usepackage{array,amsmath,amssymb,verbatim}
\usepackage{bm}% bold math
\usepackage{graphicx}% Include figure files
\usepackage{psfrag}% use latex text in ps figures

\newcommand{\bds}[1]{\boldsymbol{#1}}

\newcommand{\ddr}{\frac{\partial^2}{\partial r^2}}

\newcommand{\vR}{{\bds r}}
\newcommand{\vx}{{\bds x}}

\newcommand{\vp}{{\bds p}}
\newcommand{\vq}{{\bds q}}

\def\be{\begin{equation}}
\def\ee{\end{equation}}
\newcommand{\bea}{\begin{eqnarray}}
\newcommand{\eea}{\end{eqnarray}}
\begin{document}
\title{Dilute and Collapsed Phases of the Self-Gravitating Gas}
\author{C. Destri$^{(a)}$}
\email{Claudio.Destri@mib.infn.it}
\author{H. J. de Vega$^{(b)}$}
\email{devega@lpthe.jussieu.fr} \affiliation{
$^{(a)}$ Dipartimento di Fisica G. Occhialini, Universit\`a
Milano-Bicocca Piazza della Scienza 3, 20126 Milano and
INFN, sezione di Milano, via Celoria 16, Milano Italia\\
$^{(b)}$ LPTHE, Universit\'e Pierre~et~Marie Curie,~~Paris VI et
Denis Diderot, Paris VII, Laboratoire Associ\'e au CNRS UMR 7589,
Tour 24, 5\`eme.~~\'etage, 4, Place
Jussieu, 75252 Paris, Cedex 05, France}
\affiliation{Observatoire de Paris, LERMA.
Laboratoire Associ\'e au CNRS UMR 8112.
 \\61, Avenue de l'Observatoire, 75014 Paris, France.}
\begin{abstract}
  The self-gravitating gas in thermal equilibrium is studied using a
  Newtonian potential regularized at short distances. This short distance
  cutoff permits us to obtain a complete description of the gas including
  its collapsed phase. We give a field theory description of the $N$-body
  regularized self-gravitating gas in the canonical ensemble.  The
  corresponding functional integral is dominated in the $ N \to \infty $
  limit by saddle points which provide a mean field description. The
  well-known dilute solutions (isothermal spheres) are recovered. We find
  new solutions which are regular in the regularized theory but become
  singular in the zero cutoff limit. They describe collapsed configurations
  where the particles are densely concentrated in a region of the size of
  the cutoff. These collapsed solutions provide the absolute minimum of
  the free energy. We find further new solutions which interpolate
  between the collapsed and the dilute configurations and describe
  tunneling processes where the gas collapses. The transition probability
  for such collapse processes turns out to be extremely small for large
  $N$.  That is, the dilute solutions are in practice stable in the regime
  where they are locally stable.
\end{abstract}
\date{\today}
\maketitle 
\tableofcontents

\section{Introduction}
Selfgravitating system are ubiquous in the universe. Selfgravity governs
the formation and evolution of the large scale structure in the universe
as well as the dynamics of galaxy formation,
stars and interstellar medium \cite{astro,ism}.
It is therefore of great interest to study self-gravitating fluids in
thermal equilibrium although most astrophysical and cosmological
systems are not in exact thermal equilibrium \cite{astro}-\cite{nos}. 

The self-gravitating gas has peculiar properties from the point of view of
equilibrium statistical mechanics: the most important configurations are
inhomogeneous and the thermodynamic functions {\bf exist} in the {\bf
  dilute} limit \cite{I,II,pal}
\begin{equation}\label{limter}
  N\to \infty\; ,\quad V \to \infty\; ,\quad
  \frac{N}{V^{1/3}} \;\mbox{fixed} \; ,
\end{equation}
where $N$ is the number of particles and $ V $ stands for the volume of the
box containing them. In such a limit, the internal energy, the free energy
and the entropy turns to be extensive. That is, they take the form of $ N $
times a function of intensive dimensionless variables. In the
case of a spherical box of radius $R$ (so that $V=4\pi R^3/3$) there is
only one such variable, that is the ratio of the characteristic
gravitational energy $G m^2 N/R$ ($m$ is the particle mass and $G$
is Newton's constant) and the kinetic energy of the order of the temperature
$ T $ of a particle in the gas \cite{I,II,pal}:
\begin{equation*}
\eta = \frac{G \, m^2 N}{R \; T}  \;.
\end{equation*}
For more complex geometries there will be, besides $\eta$ (in such cases
$R$ is to be identified with the linear scale of the box), other
dimensionless shape parameters.

In this paper we restrict the analysis to the simplest case of a spherical
box and follow the approach of refs.\cite{I,II,pal} where the starting
point is the partition function for non-relativistic particles interacting
through their gravitational attraction in thermal equilibrium. The
configurational part of this partition function is expressed as a
functional integral over a scalar field $\phi(\vx)$ [proportional to the
the gravitational potential] with an Euclidean action (that is minus the
logarithm of statistical weight) proportional to $N$. Therefore, the saddle
point technique (or mean field method) permits to compute the configuration
partition function in the large $ N $ limit.

As investigated in refs.\cite{viej,I,II,pal} the self-gravitating gas in the
canonical ensemble is in a gaseous phase for $ 0 \leq \eta < \eta_T $ where
$ \eta_T = 2.43450\ldots $. (Further studies on self-gravitating gases are reported
in ref. \cite{viej2,nos,otros,cha}). At this point the isothermal compressibility
diverges and the speed of sound inside the sphere becomes imaginary.
This clearly shows that the gas phase collapses at this point in the
canonical
ensemble as a consequence of the Jeans instability. Monte Carlo
simulations explicitly show that the gas collapses for $ \eta \gtrsim
\eta_T $ into a very dense and compact phase\cite{I}.

It must be recalled that the small fluctuations around the mean field
solution in the canonical ensemble are stable for  $ \eta < \eta_C = 2.517551
\ldots $\cite{II}. Namely, fluctuations in the partition function do not
feel the divergence of the isothermal compressibility neither
the fact that the  speed of sound becomes imaginary.

Even in the regime where the gaseous phase is locally stable
there are always collapsed configurations which have much
lower energy than the gas. In fact, the energy has its absolute minimum in
the collapsed phase where all particles fall to one point and this
configuration completely dominates the free energy, causing its divergence
to minus infinity. Of course, interactions other than gravitational always
dominate for short distances. Therefore, the properties the collapsed phase
depend on the short distance physics. This physics is different according
to the nature of the `particles' considered. They can be from galaxies till
molecules or atoms. Still, we can study the regime when collapse
happens just introducing a short distance cutoff (ultraviolet regulator)
to the gravitational interaction.

We present a complete description of the self-gravitating
gas, including the collapsed phase, using a Newtonian potential regularized
at short distances. We choose the Newton potential to be $ -G \; m^2 / A $
for interparticle distances $ r \leq A $ instead of Newton's law
 $ -G \; m^2 / r $. The cutoff $ A $ is chosen very small compared with the 
size of the box.

We give an exact mapping of the $N$-body regularized problem into a field
theory functional integral. This functional integral is dominated
by its saddle points in the $ N \to \infty $, namely by the mean field
equations. The mean field equations coincide for zero cutoff with
the Lane-Emden equation obtained from hydrostatics and assuming an ideal gas
equation of state.

In this way we reproduce the well-known isothermal sphere solutions
describing a self-gravitating gas in local thermal equilibrium.
These  diluted solutions get small corrections for small cutoff.

We find a host of new solutions (saddle points) from our regularized
mean field equations. These new solutions are singular in the limit
of zero cutoff and cannot be find from the standard (no cutoff)
Lane-Emden equation. 
In these new solutions, which are perfectly regular for non-zero cutoff,
a large number of particles concentrate near the origin.
There are two new types of solutions.

The first type are the collapsed saddle point solutions: all the
particles are densely concentrated in a region of size of
the order of the cutoff. These solutions are studied analytically
and provide the absolute minima of the free energy, which reads,
at dominant order for small cutoff $ A \ll R $:
\begin{equation*}
  F-F_0 \simeq
  -\frac{G \; m^2 \, N(N-1)}{2\,A} +NT\,\log\frac{R^3}{(A/2)^3}\;.
\end{equation*}
Here, $ T $ is the temperature and $ F_0 $ is the free energy of an ideal gas.
The first term is just the potential energy of $ N $ particles clustered in
a small sphere of radius $ A/2 $, where the regularized gravitational
interaction is the same for all $\frac12 N(N-1)$ particle pairs. The second
term is $ T $ times the entropy loss in the collapse.  This free energy is
large and negative, unbounded from below as $ A\to0 $. In particular, this
free energy is well below the free energy of the dilute solution.

The second type of saddle point solutions interpolate between the gas phase
and the collapsed phase, in a sort of pre--collapse.  This solution has a
finite action in the zero cutoff limit and describes the tunneling
transition where a small fraction of particles coalesce into a region of
the size of the short--distance regulator so that the density near the
origin is very large.  The tunneling probability to collapse for a dilute
solution turns out to be extremely small for large $N$ and small $A$. This
means in practice that dilute solutions which are locally stable can be
considered stable. As mentioned above this happens in the canonical
ensemble for $ 0 \leq \eta < \eta_T $ where $ \eta_T = 2.43450\ldots $.

The parameter $ \eta $ for the collapsed solutions can be arbitrarily large.
That is, the particle density does not need to be dilute $ N \sim
V^{\frac13} $ [see eq.(\ref{limter})] but we can have $ N \sim V $.

We obtain in sec. \ref{metaB} an estimate for the
lifetime for the dilute phase of the self-gravitating gas:
\begin{equation*}
  \tau \sim \frac1{a}\,\sqrt{\log \frac1{a}} \; 
  e^{\frac{9 \, N}{2\,\eta}\,a\,(\log   a)^2} \sim 
  \frac{R}{A} \, \sqrt{\log \frac{R}{A}} \; 
  \exp\Big[\frac{9\,A\,T}{2\,G\,m^2}\,
  \Big(\log\frac{R}A\Big)^2 \Big] \; .
\end{equation*}
One can see that the lifetime becomes infinitely long in the zero cutoff
limit as well as when $ N\to\infty $ at fixed cutoff
[recall that $ R\sim N $ in the dilute limit of eq.~(\ref{limter})].

In summary, we provide through mean field theory a complete statistical
description of the self-gravitating gas including the absolute minimum
of the free energy.

\section{Mean Field Approach to the Self-Gravitating Gas}

At short distances, the particle interaction for the self-gravitating gas
in physical situations is not gravitational. Its exact nature depends on
the problem under consideration (opacity limit, Van der Waals forces for
molecules etc.).  We shall just assume a null short--distance pair force;
that is, we consider the Hamiltonian
\begin{equation*}
  H_N = \sum_{j=1}^N \frac{p^2_j}{2m} - 
\sum_{j<k}\frac{G \, m^2}{|\vq_j-\vq_k|_A} \quad ,
\end{equation*}
where
\begin{equation}\label{defva}
  |\vq|_A = 
  \begin{cases}\displaystyle
    |\vq| \quad, & \mbox{for} \; |\vq|\ge A \\ 
    A \quad\,, &\mbox{for}\; |\vq| \le A \quad ,
  \end{cases}
\end{equation}
and $ A \ll R $ is the short--distance cut-off. 

It must be stressed that the results presented in this work using
the soft-core cut-off eq.(\ref{defva}) will be qualitatively the same
for other types of cut-off since $ R \gg A$.

The partition function in the canonical ensemble reads
\begin{equation}\label{fp}
  Z_N = \frac1{N!\,\hslash^{3N}}\int \frac{d^{3N}\!p}{(2\pi)^{3N}} 
  \int_{V^N} d^{3N}\!q \;e^{-\beta H_N} \; ,
\end{equation}
where $ T \equiv \beta^{-1} $ is the temperature. Computing the integrals 
over the momenta $\vp_j, \; (1 \leq j \leq N) $,
\begin{equation*}
  \int_{-\infty}^{+\infty}\;\frac{d^3p}{(2\pi)^3}\; \exp\Big[- \frac{\beta
      p^2}{2m}\Big] = \left(\frac{m}{2\pi \beta}\right)^{3/2} \; ,
\end{equation*}
yields
\begin{equation}\label{fpconf}
   Z_N = Z_N^{(0)}\,  Z_N^{\rm conf} \quad,\qquad
  Z_N^{\rm conf} = \frac1{V^N} \int_{V^N} d^{3N}\!q   \prod_{1\leq j < k\leq N}
   \exp\left(\frac{G\, m^2\, \beta}{|\vq_j-\vq_k|_A}\right) \; ,
\end{equation}
where 
\begin{equation*}
   Z_N^{(0)} =\left(\frac{2\pi m}{\beta\hbar^2}\right)^{3N/2}
   \,\frac{V^N}{N!} \equiv \exp\left[-\beta \,F^{(0)}\right] 
 \end{equation*}
 is the partition function of the ideal gas. Thus,
\begin{equation*}
  F=F^{(0)}-T \; \log Z_N^{\rm conf}
\end{equation*}
is the full canonical free energy.

We can now replace the integration over the particle coordinates with a
functional integration over the configurations of the gravitational field
$\varphi$ produced by the particles. We start from the fundamental property
of Gaussian functional integrals, namely
\begin{equation}\label{gauso}
  \int{\cal D}\varphi\, \exp\left\{- S[\varphi] + \int d^3x\,
  J(\vx)\,\varphi(\vx)\right\} = \exp\left[\frac12\int d^3x
  \int d^3x'\,J(\vx)\; {\cal G}(\vx,\vx')\; J(\vx')\right] \; ,
\end{equation}
where $S[\varphi]$ is the quadratic action functional
\begin{equation*}
  S[\varphi] = \frac12\int d^3x\int d^3x'\; \varphi(\vx)\,
  {\cal G}^{-1}(\vx,\vx')\, \varphi(\vx')
\end{equation*}
and the integration measure is assumed to be normalized such that 
$\int{\cal D}\varphi\,\exp\{-S[\varphi]\}=1$. Then we set 
\begin{equation}\label{potG}
  {\cal G}(\vx,\vx') = \frac{G\, \beta^{-1}}{|\vx-\vx'|_A}
\end{equation}
and we identify the source field $J(\vx)$ as $\beta m$ times the
microscopic particle density
\begin{equation}\label{densu}
  J(\vx) = \beta m\sum_{j=1}^N \delta^{(3)}(\vx-\vq_j) \; .
\end{equation}
Inserting eqs.~(\ref{potG}) and (\ref{densu}) into eq.~(\ref{gauso}) yields,
\begin{equation}\label{idu}
  \int{\cal D}\varphi\, \exp\Bigg\{-S[\varphi] +\beta \;  m\sum_{j=1}^N 
    \varphi(\vq_j)\Bigg\} = \exp\Bigg[\frac12\sum_{j,k=1}^N 
    \frac{G \, m^2 \, \beta}{|\vq_j-\vq_k|_A}\Bigg] =
    e^{\frac{Gm^2\beta\,N}{2A}} \!\!\prod_{1\leq j<k\leq N}
   \exp\left(\frac{G \; m^2 \; \beta}{|\vq_j-\vq_k|_A}\right) \; .
 \end{equation}
This provides a functional integral representation of the product over 
particle pairs in the integrand of eq.~(\ref{fpconf}). 
Inserting the identity eq.~(\ref{idu})
in eq.~(\ref{fpconf}) and exchanging the functional integration with the
integration over the coordinates $\vq_j$ we obtain
\begin{equation}
  Z_N^{\rm conf} =  e^{-\frac{Gm^2\beta\,N}{2A}}
  \int D\varphi \; \exp\left\{-S[\varphi]\right\} \;
  \frac1{V^N} \int_{V^N} d^{3N}\!q
  \; \exp\Big[\beta \,  m\sum_{j=1}^{N} \varphi({\vq}_j) \Big] \; .
\end{equation}
where
\begin{equation*}
  S[\varphi] = -\frac{\beta}{8 \, \pi \,  G} \int d^3x \; 
  \varphi \, \nabla^2_A\varphi
\end{equation*}
and $ \nabla^2_A=-(4\pi \, G/\beta) \; {\cal G}^{-1} $ satisfies
\begin{equation*}
  -\nabla^2_A \frac{1}{|\vx-\vx'|_A} = 4\pi\,\delta^{(3)}(\vx -\vx') \; .
\end{equation*}
Since the $ {\vq}_j $ are dummy variables, we have
\begin{equation*}
    \int_{V^N} d^{3N}\!q  \; \exp\Big[\beta \, m\sum_{l=1}^{N} 
    \varphi({\vq}_l) \Big]= \left[ \int_{V}  d^3q \; 
      e^{\beta \,  m \, \varphi({\vq})} \right]^N 
\end{equation*}
and the configuration partition function becomes
\begin{equation}\label{rizc}
  Z_N^{\rm conf} = e^{-\frac{G \, m^2 \, \beta\,N}{2 \, A}}
  \int{\cal D}\varphi\, e^{-S_{\rm eff}[\varphi]} \; ,
\end{equation}
where
\begin{equation}\label{accef}
  S_{\rm eff}[\varphi] = S[\varphi] + S_{\rm int}[\varphi] \quad,\qquad 
  S_{\rm int}[\varphi] \equiv - N\log\Big[\frac1{V} \int_V d^3x \;
  e^{-\beta \, m \, \varphi(\vx)}\Big]
\end{equation}
We want to stress that this field--theoretic functional formulation of the
self-gravitating gas is fully equivalent, for any $ N $, to the original one
in terms of particles. The functional integral representation of the
partition function presented in ref.\cite{I,II} for the canonical ensemble
is slightly different but equivalent to eqs.~(\ref{rizc}) and~(\ref{accef})
for large $ N $. Let us also observe that the nonlocal operator $ \nabla^2_A $
reduces to the standard Laplacian operator in the limit $A\to 0$. However,
this limit is not so straightforward in the functional integral above,
since the interaction term $ S_{\rm int} $ in the action may (and indeed
does) become unbounded from below in such limit.

We now pass to dimensionless variables by setting $ \vR=\vx/R $ and 
\begin{equation*}
  -\beta m\, \varphi(\vx) = \phi(\vR) \; .
\end{equation*}
The action for $ \phi(\vx) $ now reads
\begin{equation}\label{accefa}
   S_{\rm eff}[\varphi] = N \,s[\phi]\quad,\qquad 
   s[\phi] = -\frac1{8\pi\,\eta}\int d^3r\; \phi  \; \nabla^2_a \phi 
   - \log\left[ \int_{|\vR|\le 1} d^3r \; e^{\phi}\right] +
   \log\frac{4 \, \pi}3 \; ,
\end{equation}
where
\begin{equation}\label{defeta}
  \eta \equiv \frac{G \, m^2\beta\, N}{R} \quad,\qquad a\equiv\frac{A}R
  \quad,\qquad  -\nabla^2_a \frac{1}{|\vR-\vR'|_a} = 
  4\pi\,\delta^{(3)}(\vR -\vR') \; .
\end{equation}
We recall that the variable $ \eta $ is the ratio of the characteristic
gravitational energy $ \frac{G \,  m^2 \, N}{R} $ and the kinetic energy 
$\sim T=\beta^{-1} $ of a particle in the gas. For $\eta=0$ the ideal gas is
recovered.

Having factored out of the action the number of particles $N$, it is
natural to use the saddle points method to evaluate the functional
integral over $\phi$ in the limit $N\to\infty$ at fixed $\eta$, that is in
the dilute limit. Strictly speaking, however, the action per particle
$s[\phi]$ still depends on $N$ through $a=A/R\propto A/N$. Thus we
should better regard the saddle point method as a mean field type
approximation yielding the free energy $F$ to leading order in $N$,
as we shall now show. The stationarity condition 
\begin{equation*}
  \frac{\delta s[\phi]}{\delta \phi(\vR)} = 0 \; .
\end{equation*}
leads to the regularized self--consistent Boltzmann--Poisson 
equation
\begin{eqnarray}  \label{eq;BPLE}
&&  \nabla^2_a \phi + \frac{4\pi\,\eta}{Q}\; e^{\phi} = 0 \quad,\qquad 
  Q \equiv \int_{|\vR|\le 1} d^3r \; e^{\phi(\vR)}\quad,\qquad 
  {\rm for} \quad  | \vR| \le 1 \; , \quad {\rm and} \cr \cr
&& \nabla^2_a \phi = 0 \quad,\qquad {\rm for} \quad | \vR | \ge 1  \quad.
\end{eqnarray}
In the limit $a\to0$ this becomes the standard self--consistent
Boltzmann--Poisson (or Lane--Emden) equation derivable from hydrostatic
plus the assumption of a local, ideal gas equation of state \cite{viej}.

We may rewrite eq.~(\ref{eq;BPLE}) in integral form as
\begin{equation}\label{ecinta}
  \phi(\vR) = \eta \int d^3r' \;
  \frac{\rho(\vR')}{| \vR - \vR'|_a } \;,
\end{equation}
where 
\begin{equation}\label{ro}
  \rho(\vR) = \frac{e^{\phi(\vR)}}Q \quad {\rm for}\;|\vR|\le 1 \quad , 
  \qquad \rho(\vR) =0 \quad {\rm for} \quad |\vR| > 1 \quad , \qquad
  \int d^3r \; \rho(\vR) = 1 \; ,
\end{equation}
is the normalized particle density associated to $\phi(\vR)$.

We may also use the identity involving the standard Laplacian $\nabla^2$,
\begin{equation*}
-\nabla^2  \frac1{|\vR - \vR '|_a } =\frac1{a^2} \; \delta(|\vR-\vR'|-a) \; ,
\end{equation*}
to cast the field equation in the integro--differential form
\begin{equation}\label{eq:reg0}
  \nabla^2\phi(\vR) = -\frac{\eta}{Q \; a^2} \int_{|\vR'|\le 1} d^3R'\;
  \delta(|\vR-\vR'|-a) \; e^{\phi(\vR')} \; .
\end{equation}
Notice that the laplacian in the origin is always regular for nonzero cutoff
\begin{equation}\label{eq:r=0}
   \nabla^2\phi(0) = -\frac{\eta}{Q \; a^2} \int_{|\vR'|\le 1} d^3r' \;
   \delta(|\vR'|-a) \; e^{\phi(\vR')} \; ,
\end{equation}
implying, together with the finiteness of $ \phi(0) $, that
$\nabla\phi=0$ in the origin.

\medskip

We can recast the action of the saddle point eq.(\ref{accefa}) using the 
regularized equation of motion (\ref{eq;BPLE}) as follows,
\begin{equation}\label{acci}
s[\phi] = \frac1{2 \; Q} \; \int_{|\vR|\le 1} d^3r \; \phi(\vR) \; 
e^{\phi(\vR)} - \log\frac{3 \; Q}{4 \, \pi} \quad .
\end{equation}
The free energy for large $ N $ can be written in terms of the stationary
action as\cite{I}
\begin{equation}\label{Fpv}
  F = F_0 + N\,\frac{G\,m^2}{2A} + N\, T\, s(\eta,a) + 
  {\cal O}(1) \quad ({\rm at \; fixed}\;a) \, \; ,
\end{equation}
where $ s(\eta,a) \equiv s[\phi_{s}] $, $\phi_{s}$ is a solution of
eq.~(\ref{eq;BPLE}), and $ F_0 $ is the free energy for the ideal gas.
Notice that, since $ a $ vanishes as $ N^{-1} $ as $ N\to\infty $ 
{\em at fixed} $ A $ {\em and} $ \eta $, we can regard $ N\,T\,s(\eta,a) $ 
as extensive in the particle number, that is linear in $ N $, 
only if it is regular as $ a \to 0 $. This holds true for 
stationary point which are regular at the $ a = 0 $ limit.
We show below that there are saddle points which become singular
in the $ a \to 0 $ limit and consequently the free energy is not proportional
to $ N $. Similarly, the subleading terms in eq.~(\ref{Fpv}) are
order one only for  saddle points which  are regular for $ a = 0 $.
What is always true is that they are indeed subleading for any given saddle
point.

\section{Spherically symmetric solutions}

In the case of spherically symmetry the integration over angles
in eq.~(\ref{eq:reg0}) can be performed explicitly, yielding the
one--dimensional non--linear integro--differential equation
\begin{equation}\label{ecdar}
   \ddr[r\phi(r)] = -\frac{2 \, \pi \, \eta}{Q \, a}
   \int_{|r-a|}^{r+a} dr'\; r' \, e^{\phi(r')} \; ,
\end{equation}
where now
\begin{equation} \label{qr}
  Q = 4 \pi \int_0^1 dr\; r^2 \, e^{\phi(r)} \; .
\end{equation}
Eq.~\ref{ecdar} is to be supplemented with the boundary conditions of
smooth joining (continuity of $\phi(r)$ and $\phi'(r)$ at $r=1+a$)
with the external monopole solution $\phi(r)=\eta/r$.

Integration over the angles in eq.(\ref{ecinta}) yields the non--linear
integral equation 
\begin{equation}\label{ecinar}
  \begin{split}
    \phi(r) \,=\;& \frac{4 \, \pi \; \eta}{Q} \left\{ \frac1{{\rm max}(r,a)}
      \int_0^{|r-a|}  dr' \; r'^2 \;e^{\phi(r')}\; + \; \theta(1-r-a)
      \int_{r+a}^1 dr' \; r' \;e^{\phi(r')} \right.\\
  &\left. -\,\frac1{2\,r} \int_{|r-a|}^{r+a}  dr'  \;r' \left[
      \frac{a}2 - r -r' - \frac{(r-r')^2}{2 \,a } \right] e^{\phi(r')}\;
    \theta(1-r')\right\} \; ,
\end{split}
\end{equation}
which by itself determines $ \phi(r) $ for all values of $ r $. In particular
eq.~(\ref{ecinar}) implies that any solution is finite in the origin, since
\begin{equation}\label{ecinar0}
  \phi(0) = \frac{4 \, \pi \; \eta}{Q} \Big[ \frac1{a}\int_0^a  dr \;
  r^2 \;e^{\phi(r)}\; + \;\int_a^1 dr \; r \;e^{\phi(r)} \Big] \; .
\end{equation}
Moreover, eq.~(\ref{ecdar}) or eq.~(\ref{ecinar}) imply
\begin{equation}\label{eq:phip0}
  \phi'(0) = 0 \quad\mathrm{and} \quad \phi'(r) < 0  
  \quad \mathrm{for} \quad r > 0\; ,
\end{equation}
so that the density is monotonically decreasing away from the origin.
The simplest way to prove that $\phi'(0) \le 0$ is to multiply both sides of 
eq.~(\ref{ecdar}) by $r$ and integrate from 0 to an arbitrary value.
This yields
\begin{equation*}
  \phi'(r) = -\frac{2 \, \pi \,\eta}{Q \,a\,r^2} \int_0^r dr'\; r'
   \int_{|r'-a|}^{r'+a} dr''\; r'' \; e^{\phi(r'')} \; ,
\end{equation*}
which has a negative definite r.h.s. which vanishes when $r\to0$ as long as
$a>0$. 

The action for spherically symmetric solution becomes from
eq.(\ref{acci}),
\begin{equation}\label{accir}
s[\phi] = \frac{2 \, \pi}{Q} \; \int_0^1 r^2 \;  dr \; \phi(r) \; 
e^{\phi(r)} - \log\frac{3 \; Q}{4 \, \pi} \quad .
\end{equation}
\section{Solutions Regular as $ A \to 0 $}

The saddle point equation (\ref{eq;BPLE}) admits solutions which are
regular as $ a\to0 $ and reproduce in this limit the solutions of the standard
Lane--Emden equation. We call them dilute solutions.
These are known since longtime in the spherically
symmetric case \cite{viej}. To study such solutions we can set the cutoff $
A $ to zero from the start. Equivalently, one can take the limit $a\to0$ in
eq.~(\ref{ecdar}) with the assumption that $\phi(r)$ stays finite in the
limit for any $r$, including the origin $r=0$, and that $\phi'(0)=0$.

Assuming spherical symmetry we get $ \phi(r)=\eta/r $ 
for $ r \ge 1 $, while for  $r \le 1$ we have
\begin{equation}\label{eqrad}
  \phi''(r) + \frac2r \; \phi'(r) + \frac{4\pi\eta}{Q} \; e^{\phi(r)}= 0
  \quad,\qquad \phi'(0) = 0 \; ,
\end{equation}
with the boundary conditions
\begin{equation}\label{eq:bc}
  \phi(1) = \eta \quad,\qquad \phi'(1) = -\eta \; .
\end{equation}
Clearly there exists one and only one solution for a given $Q$, so the
classification of all solutions is equivalent to the determination of all
the allowed values of $ Q $.

We set as usual\cite{I}
\begin{equation}\label{eq:phichi}
  \phi(r) = \phi(0)+ \chi(\lambda r) \; ,\quad \phi(0) =
  \log\frac{Q \,\lambda^2}{4\pi \,\eta} \;,\quad \chi(0)=0 \; .
\end{equation}
where $ \chi(z) $ must satisfy, upon inserting eq.~(\ref{eq:phichi})
in eq.~(\ref{eqrad}),
\begin{equation}   \label{eq:fund}
  \chi''(z) + \frac2z  \; \chi'(z) + e^{\chi(z)} = 0 
  \quad,\qquad  \chi'(0)=0
\end{equation}
and, from the boundary conditions eq.~(\ref{eq:bc}) at $ r=1 $,
\begin{equation}\label{eq:bc2}
  \log\frac{Q  \; \lambda^2}{4 \, \pi \, \eta} + \chi(\lambda) = \eta
  \quad ,\qquad \lambda \; \chi'(\lambda) = -\eta \; .
\end{equation}
These two relations fix the two parameters $Q$ and $\lambda$ as functions
of $\eta$. In particular, we can rewrite $\phi(r)$ as
\begin{equation}\label{eq:phichi2}
  \phi(r;\eta) = \eta -\chi(\lambda(\eta)) + \chi(\lambda(\eta) r) \;.
\end{equation}
The density profile of eq.~(\ref{ro}) reads in terms of $\chi(\lambda r)$,
\begin{equation}
\label{eq:dprof}
  \rho(r) = \frac{\lambda^2}{4 \, \pi \; \eta} \; e^{\chi(\lambda r)} \; ,
\end{equation}
so that the normalization factor can also be written $ Q = e^\eta/\rho(1) $.

Due to scale--invariant behaviour of the  gravitational interaction, eq.
(\ref{eq:fund}) enjoys the following scale covariance property: if
$ \chi(z) $ is a solution, then also $ \chi_\alpha(z) \equiv
\chi(z\,e^\alpha)+2 \,\alpha $ is a solution. 
Using $ \chi_\alpha(z) $ rather than $ \chi(z) $ is compensated by the shift 
$ \lambda\to\lambda \, e^{-\alpha}$. Notice that $ \phi(r) $ is indeed defined 
as a scale transformation of $ \chi(z) $, with the scale parameter 
$ \lambda $ (not uniquely) fixed by $\lambda\, \chi'(\lambda) = -\eta$. 
As a consequence of this scale invariance
all physical quantities must be invariant under the simultaneous
replacements $ \lambda\to\lambda\, e^{-\alpha} $ and
$ \chi(\lambda)\to\chi(\lambda) + 2 \, \alpha $.

Let us observe that, if the solutions of eq.(\ref{eqrad}) or
(\ref{eq:fund}) did not fulfill $\phi'(0)=0$ or $ \chi'(0)=0$, a delta
function at the origin would appear in the right hand side and $
\chi(z)\simeq -A/z $, with $ A>0 $, when $z\to0$.  This is the only
possible singular behaviour and corresponds to a point particle with
negative mass $ - A $ in the origin.  We shall not consider such unphysical
solutions, which are ruled out when eq.(\ref{eqrad}) is considered as the
$a\to0$ limit of eq.~(\ref{ecdar}), and stick to $ \chi'(0)=0 $.

Moreover, scale invariance allows to set also $ \chi(0)=0 $.  This choice
completely fixes the scale and we have no more scale invariance left.  One
has from the second eq.~(\ref{eq:bc2}) that $ \chi'(z)<0 $ for all positive
$ z $, so that the density profile is monotonically decreasing with the
distance.

\begin{figure}[ht]
  \centering
  \psfrag{loglam}{$\log\lambda$}
  \psfrag{etavar}{$\eta$}
  \psfrag{etab}{$\eta_T$}
  \psfrag{etac}{$\eta_C$}
  \psfrag{eta2}{$\eta_2$}
  \includegraphics[width=.75\textwidth]{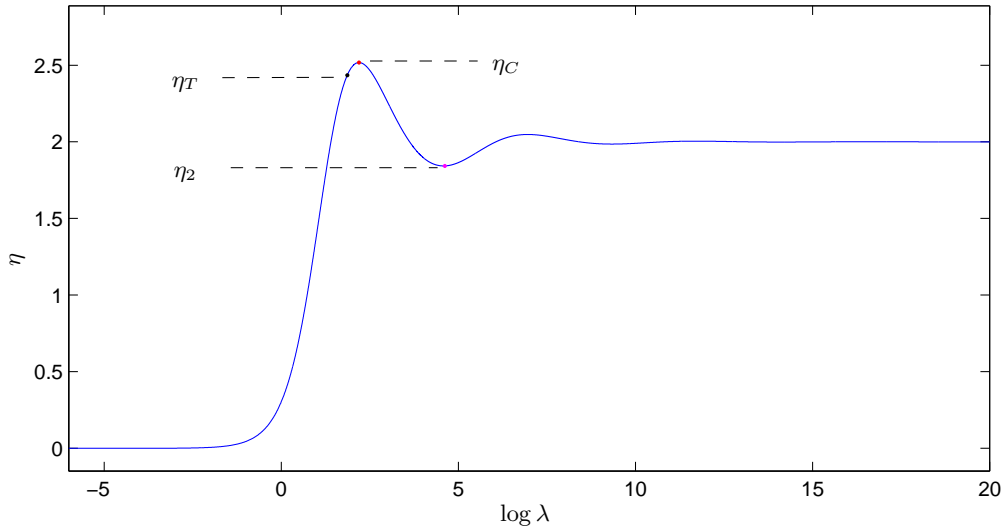}
  \caption{ $ \eta $ as a function of $ \log \lambda $ according to
    eq.(\ref{eq:bc2}) for the dilute solutions. Notice the maximum 
of $ \eta $ at $ \eta_C =
    2.51755\ldots $, where $c_V$ diverges. The region beyond the maximum
    location, at $\log \lambda_{\rm C} = 2.196459\ldots $, is locally
    unstable as discussed in ref.\cite{II}. $ c_P $ and $ \kappa_T $ diverge
    before the maximum, at the point $ \eta_T=2.43450\ldots $.}
  \label{fig:eta0} 
\end{figure}

Since $\chi(z)$ is monotonically decreasing, the allowed values of $ Q $
[that is of $ \rho(1) $, see eqs.(\ref{ro}) and (\ref{eq:bc})] 
are in one--to--one correspondence with to the
roots of the relation $ \lambda=\lambda(\eta) $ which inverts the second of
eqs.~(\ref{eq:bc2}). Thus there is only one dilute solution for each value
of $ \lambda $. However, $ \lambda=\lambda(\eta) $ is a multiple valued
function for a certain range of $\eta$. We plot $\eta $ vs. $ \log \lambda$
in Fig.~\ref{fig:eta0}.  One sees that there is a unique $\lambda $ for a
given $\eta $ only for $ \lambda< 3.6358865\ldots $, that is $ \eta
<\eta_2 \equiv 1.84273139\ldots $. For $ \eta_2<\eta<\eta_C\equiv
2.517551\ldots $ ($ \eta_C $ is the absolute maximum, located at
$\lambda_C=8.99311\ldots $, of $ -\lambda\chi'(\lambda) $ over $
0<\lambda<\infty $) the relation $ \lambda=\lambda(\eta) $ is indeed
multivalued and as $ \eta $ approches the value $ 2 $ there are increasingly
more solutions which accumulate near the purely logarithmic solution

\begin{figure}[ht]
  \centering
  \psfrag{rvar}{$r$}
  \psfrag{rho}{$\rho$}
  \psfrag{eta}{$\eta=2$}
  \psfrag{lam0}{$\lambda=4.071496\ldots$}
  \psfrag{lam1}{$\lambda=37.526798\ldots$}
  \psfrag{lam2}{$\lambda=428.912529\ldots$}
  \psfrag{lam3}{$\lambda=4532.148619\ldots$}
  \psfrag{laminf}{$\lambda=\infty$}
  \includegraphics[width=.75\textwidth]{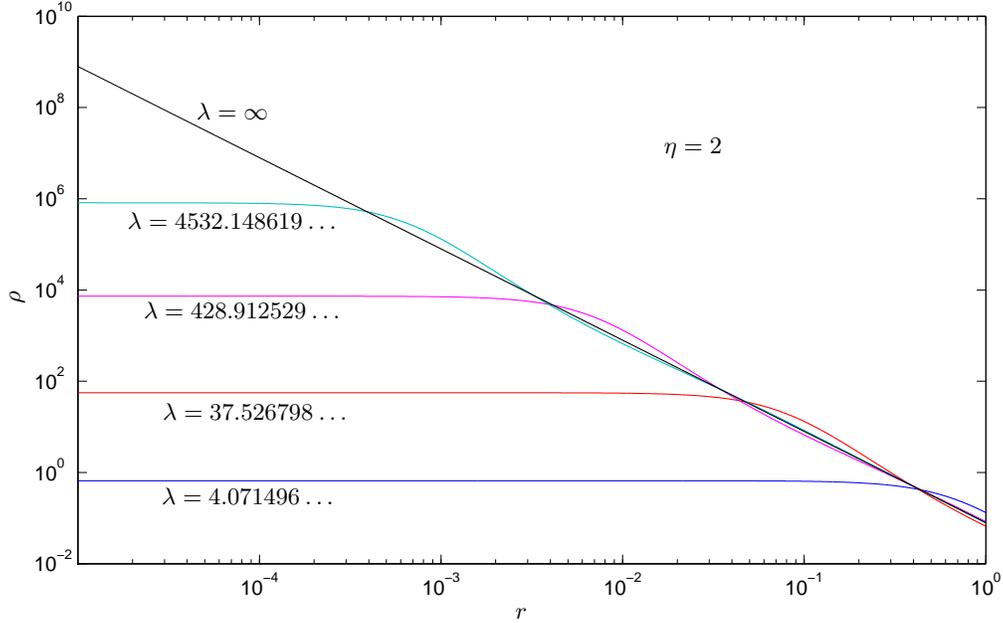}
  \caption{Log--log plot of the density profile for some dilute solutions
    at $\eta=2$. Only the smallest one at $\lambda=4.071496\ldots$ is
    locally stable.}
  \label{mrho} 
\end{figure}

\begin{equation*}
  \phi_\infty(r) = \lim_{\lambda\to\infty} \Big[\eta -\chi(\lambda) 
   + \chi(\lambda r) \Big] = 2-2\,\log r \; .
\end{equation*}
which correspond to the singular density profile $\rho_\infty(r)=(4\pi
r^2)^{-1}$. In other words, when $\eta$ is exactly 2 there exist an
infinite set of solutions corresponding to the increasing sequence of
values of $\lambda$ which satisfy $\lambda\chi'(\lambda) = -2$. In Fig.
\ref{mrho} we plot in log--log scale the density profile $\rho(r)$ [see
eq.~(\ref{eq:dprof})] vs. $r$ for few terms of this sequence.

We recall that, for any given $ \eta<\eta_C$, only the unique
dilute solution with $ \lambda<\lambda_C $ is locally stable. That is,
any small fluctuation increases the action. 
At $ \lambda=\lambda_C $ a zero--mode appears in the
linear fluctuation spectrum and $ c_V $, the specific heat at constant
volume, diverges. For larger values of $ \lambda $ negative modes show up,
signalling local instability. Actually the locally stable dilute solutions 
describe a (locally) stable dilute phase, in the thermodnamic and
mechanical terms, only for $ \eta<\eta_T=2.43450.. $. At $
\lambda=\lambda_T=6.45071\ldots $ 
the isothermal compressibility
diverges and the speed of sound inside the sphere becomes imaginary.
The gas phase collapses at this point in the canonical
ensemble as a consequence of the Jeans instability. This is confirmed by
Monte Carlo simulations\cite{I}. Hence, in the case $ \eta=2 $ depicted in
Fig. \ref{mrho}, only the solution with the smallest density contrast,
corresponding to the smallest value $ \lambda=4.071496\ldots $, is
stable. The unstable solutions that exist for $ \eta_2<\eta<\eta_T $
should imply that the dilute phase is metastable for this range of $ \eta $,
However, their lifetime is so huge [see eq.(\ref{vida2})] that these 
dilute solutions are in practice stable. 

In the stable dilute phase the particles are moderately clustered
around the origin with a density that monotonically decreases with $ r $.
One sees that the density contrast between the center and the boundary 
\begin{equation*}
  \frac{\rho(0)}{\rho(1)}= e^{-\chi(\lambda)}
\end{equation*}
grows  with  $ \lambda $ since $ \chi(\lambda) $ decreases with $ \lambda $ 
[see eq.(\ref{eq:bc2})] and Fig. \ref{mrho} \cite{viej,I,II}.

\begin{figure}[ht]
  \centering  
  \psfrag{loglam}{$\log\lambda$}
  \psfrag{lamc}{$\log\lambda_C$}
  \psfrag{etavar}{$\eta$}
  \psfrag{action}{$s$}
  \psfrag{etac}{$\eta_C$}
  \includegraphics[width=.75\textwidth]{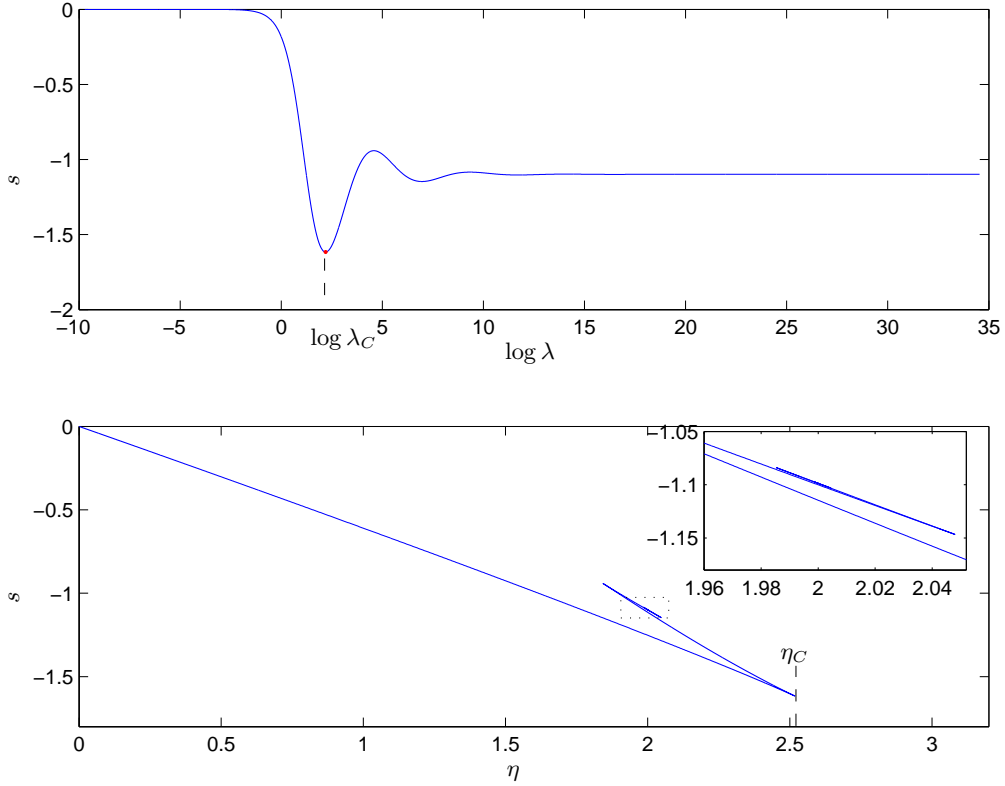}
  \caption{The action per particle $ s $ vs. $ \log\lambda $ (above) and
vs. $ \eta $ (below) for the dilute solution. $ s(\eta) $ is multivalued 
for $ \eta>\eta_2 $, with a peculiar self--similar structure, 
as evident from the inset.}
  \label{fig:action0} 
\end{figure}

The action per particle of a dilute solution can be written entirely in
terms of the density at the border\cite{I}
\begin{equation}\label{seta}
  s(\eta) = 3-\eta +\log\left[ \frac{4\pi}3 \, \rho(1)\right] -
  4\pi\rho(1)= 3-\eta +
  \log\frac{\lambda^2}{3 \, \eta}+ \chi(\lambda)-
  \frac{\lambda^2}{\eta} \, e^{\chi(\lambda)} \; ,
\end{equation}
and is multivalued for $ \eta_2 \ldots<\eta<\eta_C\ldots $. The solution
with the smallest density contrast has the smallest action, in accordance
with the linear stability analysis. We plot $ s(\eta) $ as a function of 
$\log\lambda$ and $\eta$ in Fig.~\ref{fig:action0}.

All physical quantities can be expressed  for large $ N $ 
in terms of the function
\begin{equation}\label{def}
  f(\eta) \equiv  1 + \frac{\eta}3 \; \frac{ds}{d \eta} 
  =\frac{4\pi}3\, \rho(1) \;. 
\end{equation}
We have for the energy, the pressure, the isothermal
compressibility $\kappa_T$, the specific heats ($c_V$ and $c_P$)
and the speed of sound at the boundary\cite{I}
\begin{eqnarray}\label{resto}
&&\frac{E}{3 \, N \; T} = f(\eta)-\frac12 + 
{\cal O}\left(\frac1{N}\right) \quad , \quad
\frac{p \; V}{N \; T} = f(\eta) + {\cal O}\left(\frac1{N}\right) \; , \cr \cr
&& \frac{S - S_0}{N} = -3[ 1 - f(\eta)] - s(\eta) + 
{\cal O}\left(\frac1{N}\right)
\quad , \quad \kappa_T =  \frac1{f(\eta)+\frac13 \; \eta  \; f'(\eta)} 
+ {\cal O}\left(\frac1{N}\right) \; , \\ \cr
&& c_V = 3 \left[  f(\eta)-\eta \; f'(\eta) -\frac12 \right]
+ {\cal O}\left(\frac1{N}\right) \quad , \quad
c_P = c_V +  \frac{ \left[f(\eta)-\eta \;  f'(\eta)\right]^2}{
f(\eta)+\frac13 \eta \;  f'(\eta)} + {\cal O}\left(\frac1{N}\right) \; ,\cr \cr
&&\frac{v_s^2}{T} = \frac{ \left[f(\eta)-\eta \;  f'(\eta)\right]^2}{ 3
\left[f(\eta)-\eta \;  f'(\eta)-\frac12 \right]} + 
f(\eta)+\frac13  \;  \eta \; 
f'(\eta) + {\cal O}\left(\frac1{N}\right)\; . \nonumber
\end{eqnarray}
Here $ S_0 $ stands for the entropy of the ideal gas.

Notice that $\kappa_T$ and $c_P$, the specific heat at constant pressure,
both diverge before $\eta_C$, at $ \eta= \eta_T \equiv 2.43450\ldots
$\cite{I}. Moreover, the speed of sound squared, which stays regular at the
boundary when $\eta\to\eta_T$, has a simple pole there when evaluated
inside the isothermal sphere\cite{I}. Thus, when $\eta>\eta_T$ the speed
of sound is purely imaginary inside the sphere implying that small
fluctuations grow exponentially with time.  

We have here discussed the dilute solutions in the zero cutoff limit
($A=0$). Including the short distance cutoff only adds small $ {\cal O}(A)
$ corrections to the dilute solutions.

\section{Solutions Singular as $ A \to 0 $.}

We present in this section new solutions for which the mean field $ \phi(r) $
develops singularities in the limit $ A\to0 $. In case of spherical symmetry,
when eq.(\ref{ecdar}) is the saddle point equation, the only possible 
singularity is localized at $ r=0 $. It is convenient then to `blow up'
the region near $ r=0 $ by setting [compare with eq.~(\ref{eq:phichi})]
\begin{equation}\label{eq:phixi}
  \phi(r) =  \phi(0) + \mu^2 \; \xi\left(\frac{r}{a}\right) \;, \quad 
  \phi(0)=\log\frac{Q\lambda^2}{4\pi\,\eta} \quad , \quad \xi(0) = 0  \; ,
  \quad {\rm so~that} \quad 
    \rho(r) = \frac{\lambda^2}{4 \pi \, \eta} \, e^{\mu^2 \; 
      \xi\left(\frac{r}{a}\right)}  \; .
\end{equation}
where we introduce the variable:
\begin{equation*}
\mu \equiv \lambda\,a \; .
\end{equation*}
We thus obtain for $ \xi(x) , \; x=r/a $, the following rewriting of
the integral equation (\ref{ecinar})
\begin{equation}\label{eq:varinteq}
  \begin{split}
    \xi(x) \;=\; &\frac1{{\rm max}(x,1)} \, I_2(|x-1|) - I_1(m(x))
    - \frac{(x-1)^2}{4x} \Big[I_1(m(x)) -I_1(|x-1|)\Big] \\ 
    +& \frac{x+1}{2x} \Big[I_2(m(x)) -I_2(|x-1|)\Big] - 
    \frac1{4x} \Big[I_3(m(x)) - I_3(|x-1|)\Big] - I_2(1) + I_1(1)\;,
  \end{split}
\end{equation}
where,
\begin{equation}\label{mx}
  m(x) \equiv {\rm min}\left(\frac1{a}, x+1\right) 
\;,\quad I_n(x) \equiv \int_0^{x} dy \; y^n \; \exp[\mu^2 \, \xi(y)] \; , 
\end{equation}
and we have used eq.~(\ref{ecinar0}) to write $ \phi(0) $ in terms of
$\xi(x)$ as
\begin{equation}\label{eq:phi0}
  \phi(0) = \mu^2 \left[ I_2(1) + I_1\left(\frac1{a}\right) - 
    I_1(1) \right] \; .
\end{equation}
Notice that all dependence on $ \lambda $ appears now through 
the variable $ \mu = \lambda\,a $.

\begin{figure}[ht]
  \centering  
  \psfrag{loglam}{$\log\lambda$}
  \psfrag{etavar}{$\eta$}
  \psfrag{a104}{\hspace{-0.8cm}$a=10^{-4}$}
  \psfrag{a105}{\hspace{-0.8cm}$a=10^{-5}$}
  \psfrag{a106}{\hspace{-0.8cm}$a=10^{-6}$}
  \includegraphics[width=.75\textwidth]{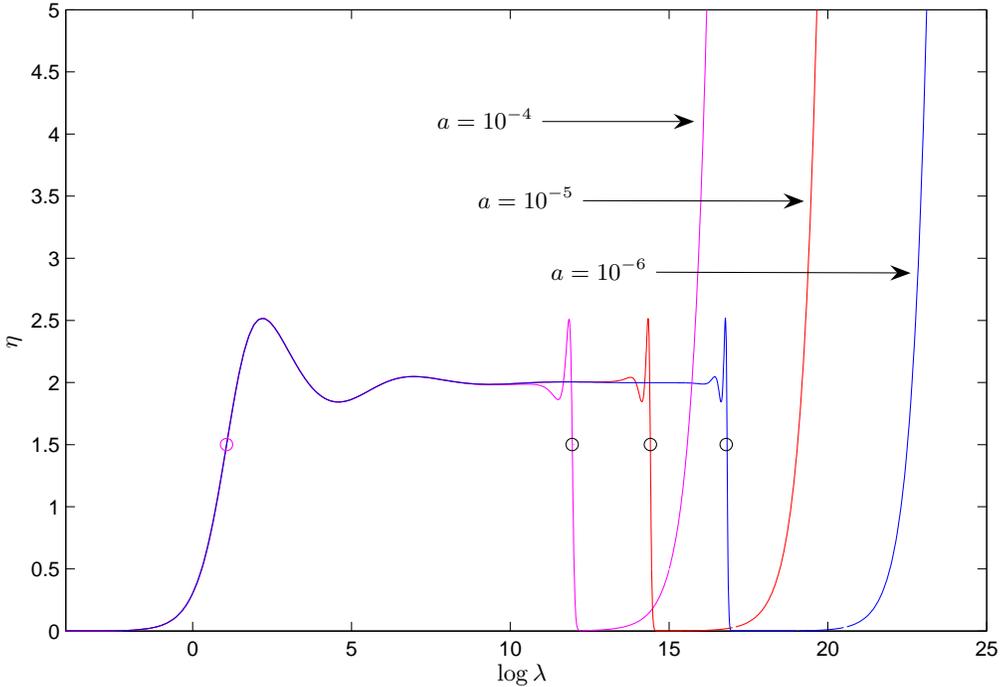}
  \caption{$ \eta $ as a function of $ \log \lambda $ according to
    eq.(\ref{eq:etamu}). The black circle mark the values of $ \lambda $
    corresponding to the interpolating solutions depicted in 
fig.~\ref{fig:chifit}. Here, $ \mu^2 \equiv \lambda^2 \; 
a^2 = {\cal O}(\log \frac1{a}) $.}  \label{fig:eta1} 
\end{figure}

By construction we have $ \xi'(0)=0 $ [see eq.~(\ref{eq:phip0})],
$ \xi(0)=0 $ and $ \xi'(x)<0 , \; \xi(x)<0 $ for all $ x>0 $. 
Notice that $ \eta $ does not enter the integral equation above, 
just as it did not
enter the differential equation (\ref{eq:fund}) for $ \chi $ in the setup
without short--distance regulator. Owing to the normalization condition
eq.~(\ref{qr}), $ \eta $ is {\bf computed} once a solution is known, 
as a function of $ \mu $ (and $ a $), through
\begin{equation}\label{eq:etamu}
  \eta = \mu^2\, a \,I_2\left(\frac1{a}\right) = 
\mu^2\, a \,\int_0^{\frac1{a}} dx\, x^2\, \exp[\mu^2\,\xi(x)] \;.
\end{equation}
>From eq.~(\ref{ecdar}) we may also derive the integro--differential form of 
the equation satisfied by $ \xi(x) $, that is 
\begin{equation}\label{eq:vardiff}
   \ddr[x \, \xi(x)] = -\frac12 \Big[I_1(x+1) -I_1(|x-1|)\Big]\;.
\end{equation}  
Eq.~(\ref{eq:varinteq}) or (\ref{eq:vardiff}) can be solved numerically to
high accuracy for any given $ \mu $. We plot the corresponding $\eta$ as a
function of $\log\lambda$ in Fig.~\ref{fig:eta1}, for better comparison
with its behaviour in the dilute case, Fig.~\ref{fig:eta0}.

We find {\bf four} different regimes 
according to the value of $ \mu $ for $ a \ll 1 $.
\begin{itemize}
\item $  \mu = {\cal O}(a) $. That is, $ \lambda = {\cal O}(1) $.
In this regime we reproduce the curve $ \eta =\eta(\lambda) $ (Fig.
\ref{fig:eta0}) characteristic of the diluted phase of sec. IV plus 
small corrections of order $ a $. 

\item $ \mu  = {\cal O}(1) $. Here $ \lambda = {\cal O}(\frac1{a}) $
and the damped oscillation pattern around $ \eta=2 $,
characteristic of the dilute solutions, gets disrupted:
the oscillations regain larger and larger amplitude until
$\mu^2=\lambda^2 a^2 \sim \log \frac1{a} $.

\item $ \mu^2  = {\cal O}(\log \frac1{a}) $. Here, a sudden drop to very small
$ \eta=\eta_{\rm min}(a)\sim a $ takes place [see fig. \ref{fig:eta1}]; 
then $ \eta $ rises again 
as $ \mu \; a $, for $ \mu \; a $ large enough, and keep growing 
indefinitely (see  fig. \ref{fig:eta1} and below). 
This regime corresponds to saddles interpolating
between the dilute phase and a collapsed phase (see below).

\item $ \mu^2  = {\cal O}(\frac1{a}) $. This regime describes collapsed
configurations where all particles are concentrated in a region of size 
$ \sim a $ and provide the absolute minimum of the free energy.

\end{itemize}

Notice that new maxima appear in addition to those present
for $ a = 0 $ (compare with Fig. \ref{fig:eta0}). Those new maxima
turn to be degenerate with the old ones up to small corrections of order $ a $.
In addition, $ \eta $ exhibits oscillations for  $ \mu  = {\cal O}(1) $
which are similar to those present at $ a=0 $ and $ \eta \to 2 $
but reversed and faster. We shall give below a qualitative explanation for
this peculiar behavior.

\begin{figure}[ht]
  \centering  
  \psfrag{log10r}{$\log_{10}r$}
  \psfrag{phiofr}{$\phi(r)$}
  \psfrag{eta1.5014}{$\eta = 1.5014\ldots$}
  \psfrag{azero}{$a=0$}
  \psfrag{a104}{$\;a=10^{-4}$}
  \psfrag{a105}{$\;a=10^{-5}$}
  \psfrag{a106}{\hspace{-0.8cm}$a=10^{-6}$}
  \psfrag{undiffeq}{unregularized differential equation}
  \psfrag{reginteq}{$\qquad$regularized integral equation}
 \includegraphics[width=.75\textwidth]{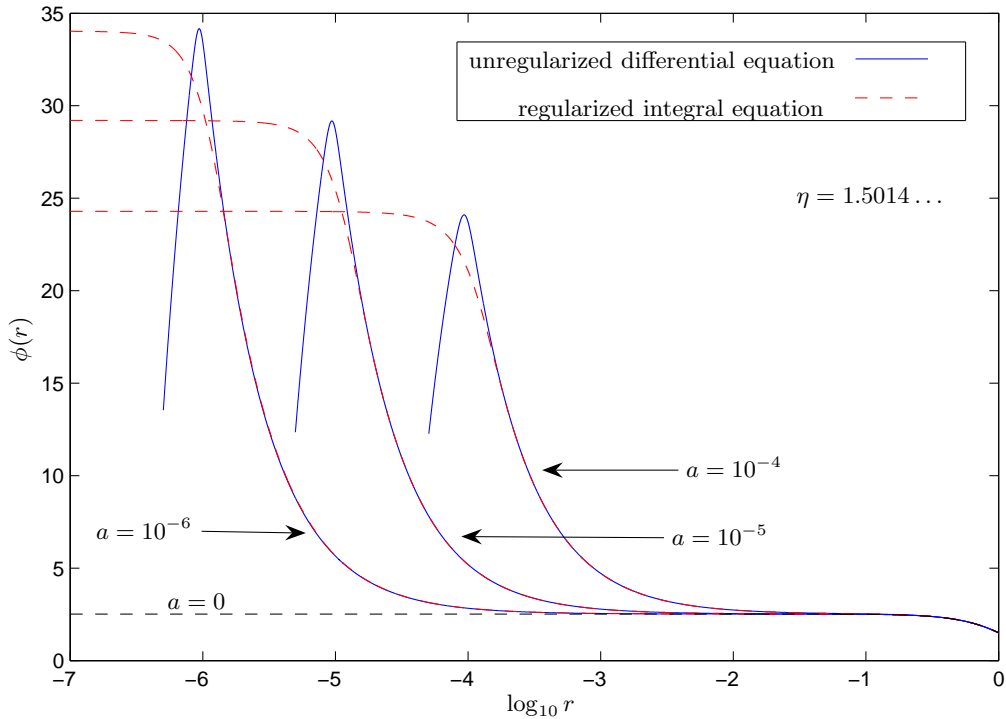}
 \caption{ Profiles of the mean field $ \phi(r) $ for $ \mu^2 = {\cal O}(\log
   \frac1{a}) $ (interpolating solutions), with comparison between the
   regularized and unregularized cases. The dashed black curve is the
   dilute solution (that is for $ a = 0 $) at the same value of $ \eta $,
   namely at the location of the magenta circle of Fig.~\ref{fig:eta1}.
The other curves correspond to the black circles in fig. \ref{fig:eta1}.}
  \label{fig:chifit} 
\end{figure}

The new dependence of $\eta$ on $\lambda$ described above implies that the
number of solutions (that is the number of distinct $ \lambda $ for any given $
\eta $) is finite for any $ \eta $ and that, most importantly, there appear
new solutions. In particular, there are two new solutions for $ \eta_{\rm
  min}(a)<\eta<\eta_2 $ in addition to the unique solution regular as $ a
\to 0 $ and there is one new solution for $ \eta > \eta_C $, the region not
accessible to dilute solutions. These new solutions are singular in the
limit $ a \to 0 $, as evident from Fig.~\ref{fig:chifit}, where we plot $
\phi(r) $ vs. $ r $ for few values of $ a $ at a fixed value of $\eta$ and
$\lambda $ placed in the sudden drop of $ \eta $ vs. $ \log\lambda $ (the
black circles in Fig.~\ref{fig:eta1}, scaling as $ \mu^2 = \lambda^2 a^2 
\sim \log \frac1{a} $).

In Fig.~\ref{fig:chifit} we also compare the solutions of the integral
equation (\ref{ecinar}) or (\ref{eq:varinteq}) to those of the differential
equation (\ref{eqrad}), with boundary conditions~(\ref{eq:bc}), in which
$Q$ takes the value obtained by the integral method, upon using
eqs.~(\ref{eq:phixi}) and~(\ref{eq:phi0}). We see that the
agreement is very good down to $ r $ of order $ a $, when the short--distance
regularization becomes effective, while the solutions of the differential
equation (\ref{eqrad}) eventually blow to $ -\infty$ as $ r\to0 $. This
agreement appears very natural from the integro--differential formulation
of eq.~(\ref{ecdar}), where the short--distance regulator $ a $ can play a
significant role only for $ r\lesssim a $.

\begin{figure}[ht]
  \centering  
  \psfrag{eta1.6108}{$\eta = 1.6108\ldots$}
  \psfrag{logr}{$\log_{10} r$}
  \psfrag{a1e-4}{$a = 10^{-4}$}
  \psfrag{logrho}{$\log \rho$}
\includegraphics[width=.75\textwidth]{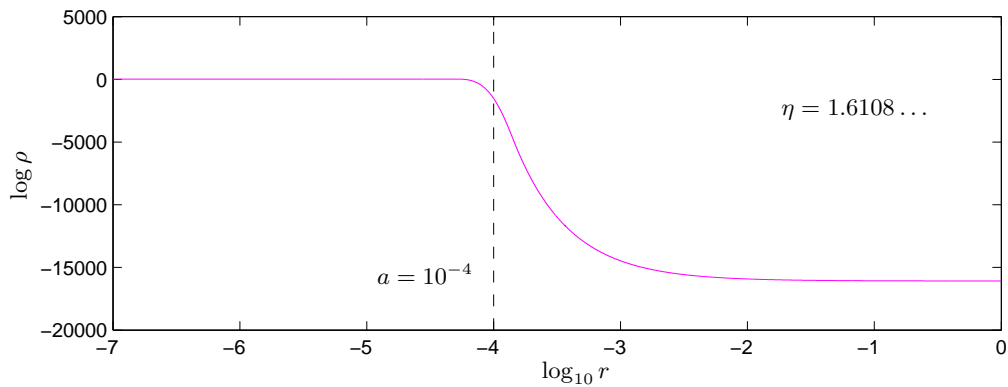}
\caption{Profiles of the logarithm the density $ \log \rho(r) $ for 
 $ \mu^2 = {\cal O}(\frac1{a}) $ (collapsed solution). For the value of
  $\eta$ given, we have $\lambda=6012149.6972\ldots$. This collpased
  profile should be compared to the dilute ones (stable and unstable)
  depicted in Fig.~\ref{mrho}. }
  \label{fig:collrho} 
\end{figure}

Hence, the qualitative behaviour of the singular solutions can be understood
already from the differential equation (\ref{eq:fund}), provided one drops
the initial conditions $ \chi(0)=\chi'(0)=0 $ characteristic of the dilute
solutions. That is, using the relation $ \chi(\mu\, x)=\mu^2 \, \,\xi(x) $
that follows from eqs. (\ref{eq:phichi}) and (\ref{eq:phixi}),
we can solve the integral equation (\ref{eq:varinteq}) just in the interval
$ 0\le x\le x_1 $, with $ x_1\gtrsim 2 $, to compute 
\begin{equation}\label{codin1}
\chi(2 \, \mu)=\mu^2\, \xi(x_1) \quad {\rm and}
\quad   \chi'(2 \, \mu)=\mu \, \xi'(x_1)  \; ,
\end{equation}
at the boundary $ x=x_1 $. These values for $ \chi(2 \,\mu) $ and $ \chi'(2
\,\mu) $, both negative by construction, can then be used as initial
conditions at $ z = 2 \, \mu $ to solve the differential equation
(\ref{eq:fund}) in place of the conditions $ \chi(0)=\chi'(0)=0 $,
characteristic of the dilute solutions. The choice $ x_1\gtrsim 2 $ is
motivated by a direct numerical analysis but can be explained by the fact
that $ \mu $ diverges when $ a\to0 $ in the solutions which are singular as
$a\to0$. This implies, on one side, that the r.h.s. of
eq.~(\ref{eq:vardiff}) is exponentially small in $ \mu^2 $ when $
x_0\gtrsim 2 $ [see also eq. (\ref{mx})]
 and, on the other side, that the initial conditions at $ z =
2 \, \mu $ affect the function $ \chi(z) $ very strongly, causing the rapid
drop for $ r > 2 \, a $ observed in Fig.~\ref{fig:chifit}, as well as the
growth followed by the fall to $ -\infty $ as $ r \to 0 $.  It must be
observed, in fact, that $ \xi(x) $ has a finite limit as $ \mu \to
\infty $ [see below eq.(\ref{scolap})], so that to leading order and 
$ \mu \gg 1 $, using eq.(\ref{codin1}), 
$$ \chi(2 \, \mu) = {\cal O} (\mu^2) \quad {\rm and } \quad 
 \chi'(2 \, \mu) = {\cal O}(\mu) \; . 
$$
It follows that the function $ \chi(z) $ fulfills approximately the
free equation $ (z\chi)''=0 $ in the rapid drop for $z>2\,\mu$ as well as in
the following plateau (to the right in Fig.~\ref{fig:chifit}), that is 
\begin{equation}\label{eq:free}
  \chi(z) \simeq \frac{c_0}z + c_1 \quad, \qquad c_0 = {\cal O}(\mu^3)\;,
  \quad c_1 = {\cal O}(\mu^2)\;. 
\end{equation}
Where we used that $ z = \lambda \; r = \mu \; x $.
This behaviour is valid as long as $ \chi''(z)\lesssim e^{\chi(z)} $, that is
$ z\lesssim {\cal O}(\mu\, \exp[\mu^2/3]) $ since $ \chi= {\cal O}(\mu^2) $
and $  c_0 = {\cal O}(\mu^3) $. For larger values of $ z $ the
function $ \chi(z) $ tends to the unique large distance fixed point given by
the purely logarithmic solution $ \log(2/z^2) $ and does this through the
characteristic damped harmonic oscillations in $\log\lambda$\cite{viej}
responsible for the waving behaviour of $ \eta $ in the dilute phase 
[see fig. \ref{fig:eta0}]. Since
the crossover to the logarithmic behaviour takes place for larger and
larger values of $ z $ the smaller is $ a $, because $ \mu = 
{\cal O}\left( \sqrt{ \log \frac1{a} } \right) $ grows with $ a $ as well
as the crossover point $ z =  {\cal O}(\mu\, \exp[\mu^2/3]) $. 
When $ \chi(z) $ is almost constant,
the characteristic oscillations tend to have the same amplitude they have
in the regular solutions. Moreover, this almost constant value of
$ \chi(z) $, which is of order $ \mu^2 $, decreases faster in $ \mu $ than the
end point $ \lambda=\mu/a $ at which $ \chi(z) $ is to be evaluated to give
$ \eta $ as in the second of eqs.~(\ref{eq:bc2}). This explain why the
oscillations of $\eta$ are reversed and faster right before the sudden drop
in Fig.~\ref{fig:eta1}. In other words, the oscillations in the left part
of fig. \ref{fig:eta1} describe the approach to the limiting
solution $ \log(2/z^2) $ [$\eta = 2$] while one is getting {\bf off} 
this limiting solution in the right part of fig. \ref{fig:eta1}.
This explains why the oscillations on the right get reversed compared
with the oscillations on the left of fig. \ref{fig:eta1}.

In the next two subsections we will provide more details on all these
matters and further analytical insights.

\subsection{Collapsed phase}

This is the case when $ \mu\sim \eta/a $ as $ a\to0 $ and corresponds to the
growing branch on the right of Fig.~\ref{fig:eta1}. Notice that we have
collapsed solutions for arbitrarily large values of $ \eta $ contrary
to the dilute solutions which only exist for $ \eta < \eta_C $.

Since $ \xi(x) $ is negative and monotonically decreasing for $x>0$, the
quantity $ \exp[\mu \, \xi(x)] $ is exponentially small except in the
interval $ 0 \le x <x_0 $ where $ \xi(x) $ would vanish in the limit. We
use now this property to get an approximate but analytic singular solution.
That is, the dominant contribution in the r.h.s. of eq.~(\ref{eq:varinteq})
is obtained when all integrations are restricted to the interval $ 0 \le x
\le x_0 $. One easily realizes that, by consistency, $ x_0=1/2 $. Then one
finds
\begin{equation*}
  I_n(x) = \frac1{n+1} 
  \begin{cases}
    x^{n+1} \;,& x \le 1/2 \\ 2^{-n-1} \; , & x \ge 1/2
  \end{cases}
\end{equation*}
and in particular using eqs.(\ref{eq:phi0}) and (\ref{eq:etamu}),
\begin{equation*}
  \phi(0) = \mu^2 \left[I_2(1) + I_1\left(\frac1{a}\right) - I_1(1)\right] = 
  \frac{\mu^2}{24} \quad,\qquad
    \eta = \mu^2\, a \,I_2\left(\frac1{a}\right) = \frac{\mu^2\,a}{24} 
\quad,\qquad \phi(0) = \frac{\eta}{a} \;.
\end{equation*}
Finally, the leading form for $\phi(r)$ reads
\begin{equation}\label{scolap}
  \phi(r) = \phi(0) + \frac{24\,\eta}a\, \xi\left(\frac{r}{a}\right) = 
  \begin{cases}\displaystyle
    \frac{\eta}a \;, & r\le a/2 \\[0.3cm]\displaystyle
    \frac{\eta}a +\frac{\eta}{2r}\left(\frac{r}a - \frac12 \right)^3 
    \left(\frac{r}a - \frac52 \right) \;, & a/2 \le r \le 3a/2 \\[0.3cm]
    \displaystyle \frac{\eta}r \;, &r \ge 3a/2 \;.  
  \end{cases}
\end{equation}
Notice that $ \phi(r) $ as well as its first and second derivatives are
continuous at $ r = a / 2 $ and $ r = 3 \, a / 2 $. For $r \ge 3a/2$, the
effective gravitational field has the free form proper of a locally
vanishing mass density.

Higher corrections can be obtained by iteration over
eq.~(\ref{eq:varinteq}). That is, using the leading form of $\xi(x)$
that can be read out of in eq.~(\ref{scolap}) to evaluate to
next--to--leading order, as $\mu\to\infty$, the integrals $I_n(x)$ in the
r.h.s. of eq.~(\ref{eq:varinteq}). This procedure can then be repeated to
obtain further corrections.

Eq.~(\ref{scolap}) provides the spherically symmetric, leading order
singular solution of the mean field equation (\ref{eq;BPLE}) where the
particles are densely concentrated around one point in a region of size of
the order of the cutoff $ A $.  In fact, to leading order in $ a $ the density 
reads
\begin{equation*}
  \rho(r) = 
  \begin{cases}\displaystyle
    \frac{6}{\pi\,a^3} \;, & r\le a/2 \\[0.3cm]\displaystyle
    0 \;, &r\ge a/2 \;,
  \end{cases}
\end{equation*}
while to next--to--leading order we have
\begin{equation}\label{eq:nnlo}
  \rho(r) = \frac{6}{\pi\,a^3} \exp\Big[\phi(r) -\frac{\eta}a \Big] \;,
\end{equation}
where $\phi(r)$ has the leading order form of eq.~(\ref{scolap}). This
corresponds to the situation when the gravitational attraction completely
overcomes the kinetic energy and all particles fall towards the origin. Of
course, such singular solution only exist mathematically for non-zero
values of the cutoff $ A $.

Let us now compute the action for the solution eq.~(\ref{scolap}) from
eq.~(\ref{accefa}). The regularized laplacian can be computed  from the
equation of motion (\ref{eq;BPLE}) with the leading order result for 
$ a\to 0 $,
\begin{equation*}
  \int_{|\vR|\le 1} d^3 r \; \phi({\bds r}) \; \nabla^2_a  \phi({\bds r}) =
  - \frac{4 \, \pi \, \eta}{Q} \; 4 \, \pi \int_0^1 r^2 \; dr \; 
  \phi(r) \, e^{\phi(r)} = - 4 \, \pi \; \eta \; \phi(0) = -\frac{4 \,
    \pi \, \eta^2}{a} \; .
\end{equation*}
The second term is just 
\begin{equation*}
  \log Q = \phi(0) - \log\frac{\lambda^2}{4\pi\,\eta} =
  \frac{\eta}a - \log\frac6{\pi\,a^3} \;,
\end{equation*}
so that all together, to leading order we obtain for the free energy
\begin{equation}\label{acsing}
  \begin{split}
    F-F_0 =\frac{Gm^2\beta\,N}{2A} + NT\,s(\eta,a) &= \frac{Gm^2\beta\,N}{2A}
    -\frac{NT\,\eta}{2\,a } + NT\,\log\frac8{a^3}\\ & =
    -\frac{Gm^2N(N-1)}{2\,A} +NT\,\log\frac{R^3}{(A/2)^3}\;.
  \end{split}
\end{equation}
The first term is just the potential energy of $ N $ particles clustered in
a small sphere of radius $A/2$, where the regularized gravitational
interaction is the same for all $\frac12 N(N-1)$ particle pairs. The second
term is $T$ times the entropy loss in the collapse.  As one could expect,
this free energy is large and negative, unbounded from below as $A\to0$. In
particular, this free energy is well below the free energy of the dilute
solution eq.~(\ref{seta}) and Fig.  \ref{fig:action0} for all values $ \eta
> 0 $.  That is, the singular solution eq.~(\ref{scolap}) provides the {\bf
  absolute} minimum for the free energy eq.~(\ref{accefa}) for any $ \eta >
0 $. The dilute solution eqs.(\ref{eq:fund}) and (\ref{eq:bc}) is only a
{\bf local} minimum. In the next subsection we shall address the issue of
the metastability of the dilute phase.

Let us now compute some physical quantities characterizing the collapsed
phase.

\begin{figure}[ht]
  \centering  
  \psfrag{etavar}{$\eta$}
  \psfrag{ffun}{$PV/NT$}
  \psfrag{  a1e-4}{$\;a=10^{-4}$}
  \psfrag{muoa}{$\mu={\cal O}(a)$}
  \psfrag{muo1}{$\mu={\cal O}(1)$}
  \psfrag{muologa}{$\mu^2={\cal O}(\log \frac1{a} )$}
  \psfrag{muo1oa}{$\mu^2={\cal O}(\frac1{a})$}
  \includegraphics[width=.75\textwidth]{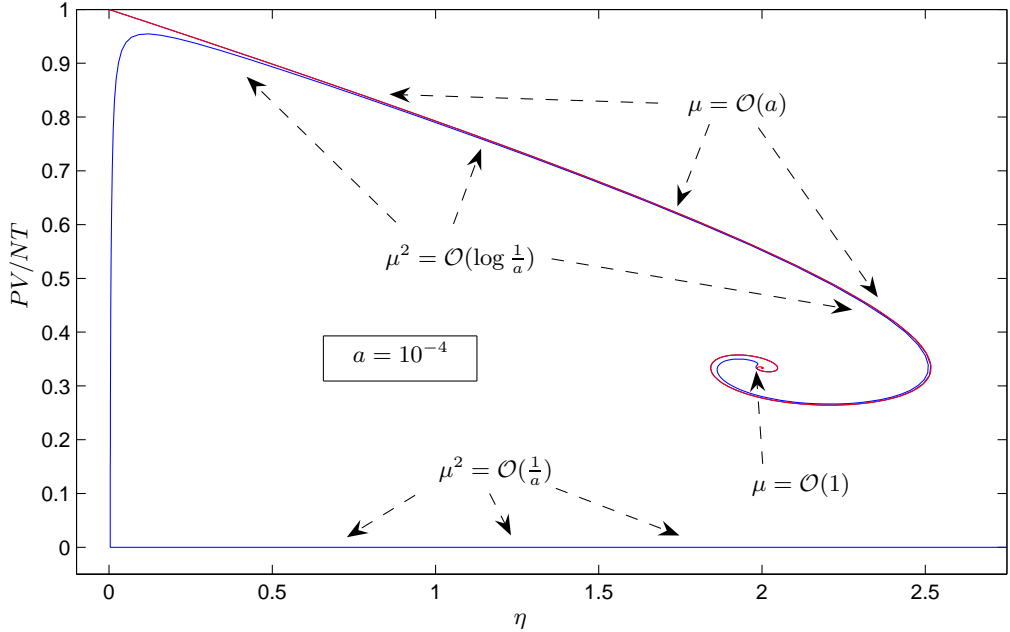}
  \caption{Complete plot of $ PV/NT $ vs. $ \eta $ when $ a=10^{-4} $ (blue
    curve). When $ \mu $ is of order $ a $ the blue curve cannot be
    distinguished from the corresponding curve of the case $ a=0 $, here
    depicted in red (dilute solutions). 
When $ \mu $ is of order $ 1 $ the blue curve ($ a>0
    $) and the red curve separate. The blue curve then traces the red
    spiral backwards at a distance of order $ a \; (\log a)^2 $ as long as
    $ \mu^2 \lesssim {\cal O}(\log \frac1{a}) $ (interpolating solutions).
 Finally the blue curve drops to zero when $ \mu^2 $ is of order $\frac1{a}$
(collapsed solutions).}
  \label{fig:ffun} 
\end{figure}

The equation of state has the mean field form
\begin{equation}\label{eq:eos}
    \frac{P \,V}{N\,T}=\frac{4\pi}3\, \rho(1) \;, 
\end{equation}
in both the dilute and collapsed phase. In the former case the r.h.s. of
eq.~(\ref{eq:eos}) coincides with the function $f(\eta)$. In the latter 
we obtain instead, from eqs.~(\ref{eq:nnlo}) and~(\ref{scolap}), to
next--to--leading order
\begin{equation}\label{eq:eos2}
\frac{P\,V}{N\,T}=\left(\frac2{a}\right)^3 \; 
\exp\Big[\eta -\frac{\eta}a \Big]  \;.
\end{equation}
Thus, the pressure at the boundary is exponentially small in the collapsed
phase. The complete plot of  $ PV/NT $ vs. $ \eta $ is given in
Fig.~\ref{fig:ffun}. One can see how the well known \cite{viej} spiraling
towards $ (\eta=2,f(\eta)=1/3) $, characteristic of the dilute solutions, is
now limited to the case when $ \mu $ is of order $ a $ . When $ \mu $ reaches
values of order $ 1 $ the inward spiral comes to a halt and then winds back
until $ \mu \lesssim {\cal O}(\log \frac1{a}) $. For larger values of $ \mu $, 
that is $ \mu\sim \eta/a $ as in the collapsed phase, $ PV/NT $ drops to values
exponentially small in $ \frac1{a} $. Analogous plots are obtained for other 
types of cutoff in refs. \cite{otros}.

For other physically relevant quantities we similarly find
\begin{eqnarray}
  c_V = \frac32 \quad , \quad c_P = \frac32 + {\cal O}(a^2)
  \quad , \quad  \frac{v_s^2}{T} = \frac1{\kappa_T} = 
  \frac8{a^3} \exp\Big[-\frac{\eta}a \Big]
  \quad , \quad \frac{S - S_0}{N} = 3 \; \log a + {\cal O}(a^0) \; .
\end{eqnarray}

\subsection{Interpolating solution: metastability}\label{metaB}

The dilute solutions studied in sec. IV are metastable because the
collapsed solution provides the absolute minimum for free energy
[eq.(\ref{acsing})]. The collapse transition from the dilute phase to the
collapsed phase can take place through an interpolating solution in an
analogous way to the bubbles in liquid-gas first order phase transitions
\cite{transi}.

Such interpolating solutions are the saddle points of eq.~(\ref{eq;BPLE}),
which correspond to the sudden drop of $ \eta $ as a function of
$\log\lambda$ (see Fig.~\ref{fig:eta1}) located approximately at $ \lambda^2
a^2\sim \log \frac1{a} $. In this narrow drop region we have 
$ {\cal O}(a) \lesssim \eta < \eta_C $, so the interpolating solutions 
are allowed only in this interval of $ \eta $. Examples of the interpolating 
solutions are given in fig.~\ref{fig:chifit}.

According to the qualitative discussion below eq.~(\ref{eq:free}), the
location of the $ \eta $ drop can be obtained approximately as the values
of $ \lambda $ at which the function $ \chi(\lambda) $ has the crossover from
the free solution $ \chi(\lambda)\simeq c_0/\lambda +c_1 $ to the logarithmic
solution $ \log(2/\lambda^2) $. That is for $ \lambda^2 a^2 \sim \log
\frac1{a} $, since $ c_0={\cal O}(\lambda^3a^3) $ and $ c_1={\cal
  O}(\lambda^2a^2) $, see eq.~(\ref{eq:free}). Then the scaling as $a\to0$
of $\rho(0)$, the density in the origin, can be obtained from
eq.~(\ref{eq:phixi}) and the numerical analysis as
\begin{equation*}
  \rho(0) = \frac{\lambda^2}{4\pi\,\eta} = 
  {\cal O}\big(a^{-2}\log\tfrac1{a}\big) \;.
\end{equation*}
because $ \eta = {\cal O}(1) $ in this regime.

On the other hand, from
Fig.~\ref{fig:chifit} one sees that the value of the density for $r\gg a$
and in particular the value at the border $ \rho(1) $ differ by order $a$
w.r.t. the dilute phase. The equation of state (\ref{eq:eos}), which holds
for any value of $ \lambda $, then implies that the pressure in the dilute
and interpolating solution, for the same value of $ \eta $, differ only by
order $ a $, as evident also from Fig.~\ref{fig:ffun}.

In practice, one can regard an interpolating configuration as a special
spherical fluctuation of the dilute phase in which a fraction of particles
of order $ a \, \log\tfrac1{a} $ is almost uniformely removed from everywhere 
and concentrated in a region of size $ a $ around the origin. In fact, this
result has an heuristic explanation by a simple energy--entropy argument: if
we denote by $ \Delta\rho $ the variation of the density in a region of size
$a$ around the origin, we then have
\begin{equation}\label{eq:enent} 
  \Delta s  \sim  -\frac\eta{2\,a} \,  \left(\frac{\Delta N}N\right)^2 
  + 3 \, \frac{\Delta N}N \,\log\frac1{a} \quad,\qquad 
  \frac{\Delta N}N \equiv \,a^3 \, \Delta\rho\ ; ,
\end{equation}
since the first term is the gain in potential energy per particle and the
second the loss of entropy per particle due to the concentration. We may
neglect the effect due to the rest of the particles if we assume that
$\frac{\Delta N}N$, the fraction of particle moved around, would vanish as
$a\to0$. Minimizing eq.~(\ref{eq:enent}) w.r.t. $\Delta\rho$ yields
\begin{equation}\label{eq:drho}
  \Delta\rho \simeq \frac3{\eta\,a ^2} \, \log\frac1{a} \quad {\rm and}
 \quad \Delta s = \frac{9 \, a}{2 \, \eta} \left(\log\frac1{a}\right)^2 \;,
\end{equation}
which is indeed consistent with the requirement that $ a^3 \; \Delta\rho $ 
vanishes as $ a\to0 $ and rather accurately approximate our numerical results.

Notice that the minimized $ \Delta s $ is to be identified with the
difference, between interpolating and dilute solution, of the action per
particle, so that 
\begin{equation}\label{ds}
\Delta s =  s_{\rm interpolating} - s_{\rm dilute} 
\simeq \frac9{2\,\eta}\,a\,(\log   a)^2 \;.
\end{equation}
This result is in very good agreement with the numerical values
obtained by solving eqs.(\ref{ecinar}) and (\ref{accir}).

According to the interpretation of the interpolating solutions as saddle
points through which the dilute phase may decay to the collapsed one, 
the lifetime of the dilute phase is given by
\begin{equation}\label{vida}
  \tau \simeq \Big(\frac{|{\rm Det}|}{{\rm Det}_0}\Big)^{1/2}
  \,e^{N \Delta s} \, \;,
\end{equation}
where $ \rm Det $ stands for the determinant of small fluctuations around
the interpolating solution and $ {\rm Det}_0 $ for that around the
dilute solution. 

This ratio of determinants can be estimated as follows: in the
interpolating profile there should be only one negative mode necessarily
localized in a region of size $ a $ around the origin, where the density is
very large, since $-\rho(r)$ plays the role of Schroedinger potential in
the eigenvalue equation for the small fluctuations. Eq.~(\ref{eq:drho})
then implies that $ \frac1{a^2}\, \log \frac1{a} $ provides the
scale of the potential felt by this negative mode. Hence the corresponding
negative eigenvalue should be of the same order, while the rest of the
spectrum should be almost unaffected, since the two density profiles differ
only by order $a$ for $r \gtrsim 2\,a$. Therefore
\begin{equation}\label{dete}
  \frac{\rm |Det|}{{\rm Det}_0}\simeq  \frac1{a^2}\, \log \frac1{a} \; .
\end{equation}
We obtain from eqs. (\ref{ds})-(\ref{dete}) to leading order the following
lifetime for the dilute phase of the self-gravitating gas:
\begin{equation}\label{vida2}
  \tau \sim \frac1{a}\,\sqrt{\log \frac1{a}} \; 
  e^{\frac{9 \, N}{2\,\eta}\,a\,(\log   a)^2} \sim 
  \frac{R}{A} \, \sqrt{\log \frac{R}{A}} \; 
  \exp\Big[\frac{9\,A\,T}{2\,G\,m^2}\,
  \Big(\log\frac{R}A\Big)^2 \Big] \;.
\end{equation}
One can see that the lifetime becomes infinitely long in the zero cutoff
limit as well as when $N\to\infty$ at fixed cutoff
[recall that $ R\sim N $ in the dilute limit of eq.~(\ref{limter})].

We want to stress that the lifetime of the dilute solutions
is extremely long for large $ N $ and small cutoff $ A $. 
This is due to the fact
that although the collapsed solution has a much lower energy than the
dilute solution, in the latter there is a huge entropy barrier against the
gathering, in a small domain of size $ \sim A $, of a number particles
large enough to start the collapse. A qualitatively similar conclusion is
reached in ref. \cite{cha}.

We shall come back for a more detailed study of this problem in a
subsequent paper.


\begin{thebibliography}{}

\bibitem{astro} J. Binney, S. Tremaine, `Galactic Dynamics',
Princeton Univ. Press, Princeton, NJ,  1987. P. J. E. Peebles,
`Principles of Physical Cosmology', Princeton Univ. Press, Princeton,
NJ,  1993. W. C. Saslaw, `Gravitational Physics of Stellar and Galactic
Systems', Cambridge Univ. Press, Cambridge 1987.

\bibitem{ism} R. B. Larson, MNRAS {\bf 194}, 809 (1981).
J. M. Scalo, in `Interstellar Processes', D.J. Hollenbach and
H.A. Thronson Eds., D. Reidel Pub. Co, p. 349 (1987).

\bibitem{viej} R. Emden, `Gaskugeln', Teubner, Leipzig und Berlin,
1907. S. Chandrasekhar, `An Introduction to
the Study of Stellar Structure', Chicago Univ. Press, 1939.
R. Ebert, Z. Astrophys. {\bf 37}, 217 (1955).
W.B.Bonnor, Mon. Not. R. astr. Soc. {\bf 116}, 351 (1956).
V. A. Antonov, Vest. Leningrad Univ. 7, 135 (1962).
D. Lynden-Bell and R Wood, Mon. Not. R. astr. Soc. {\bf 138}, 495 (1968).

\bibitem{viej2}  E. B. Aronson, C. J. Hansen, Astrophys. J.  177, 145 (1972),
B. Stahl, M. K. H. Kiessling, K. Schindler, Planet. Space Sci. 43, 271 (1995),
P. Hertel, W. Thirring, Commun. Math. Phys.  24, 22  (1971).

\bibitem{I} H. J. de Vega, N. S\'anchez, Nuclear Physics
{\bf B 625}, 409 (2002), astro-ph/0101568. 

\bibitem{II} H. J. de Vega, N. S\'anchez, Nuclear Physics
{\bf B 625}, 460 (2002), astro-ph/0101567. 

\bibitem{pal}  H. J. de Vega, N. S\'anchez, 
   `Statistical Mechanics of the self-gravitating gas: thermodynamic limit, 
phase diagrams and fractal structures', astro-ph/0505561, 9th Course of the 
International School of Astrophysics `Daniel Chalonge', 
Palermo, Italy, 7-18 September 2002, NATO ASI,  p. 291-324 in the
Proceedings, N. S\'anchez and Yu. Parijskij editors, Kluwer, 2002.

\bibitem{nos}  H. J. de Vega, J. Siebert,  Phys. Rev. {\bf E66}, 016112 (2002),
Nucl. Phys. {\bf B 707}, 529 (2005) and {\bf B 726}, 464 (2005).
H. J. de Vega, N. S\'anchez, Nucl. Phys. {\bf B711}, 604 (2005).

\bibitem{otros} P. H. Chavanis, Phys. Rev. E 65, 056123 (2002),
P. H. Chavanis, I. Ispolatov, Phys. Rev. E  66, 036109 (2002),
P. H. Chavanis, J. Sommeria, MNRAS 296, 569 (1998).
E. Follana, V. Laliena, Phys. Rev. E 61, 6270 (2000).

\bibitem{transi}   Langer J S in \textit{Fluctuations,
Instabilities and Phase Transitions} ,   Riste T,  Ed. page 19,
Plenum N.Y. (1975). Langer J S in  \textit{Solids Far
From Equilibrium} , Ed.   Godreche C, Cambridge Univ. Press
(1992), page 297, and  \textit{Systems Far From Equilibrium},
Garrido L, et.\ al.\ eds. \textit{Lect. Notes in Phys.} {\bf 132}
Springer (1975). Gunton J D,   San Miguel M and
Sahni P S  in  \textit{Phase Transitions and Critical Phenomena} ,
  Domb C and   Lebowitz J J, Eds. Vol 8, Academic Press, (1983);
  Langer JS, \textit{Acta Metall.} {\bf 21}, 1649 (1973).
D. Boyanovsky, H. J. de Vega, D. J. Schwarz,
`Phase transitions in the early and the present Universe', 
hep-ph/0602002,  to appear in Ann. Rev. Nucl. Part. Sci.

\bibitem{cha} P. H. Chavanis, A\&A 432, 117 (2005).

\end{thebibliography}
\end{document}